\documentclass[aps,twocolumn,superscriptaddress]{revtex4}
\usepackage{latexsym}
\usepackage{graphicx}
\usepackage{amsmath}

\begin{document}


\title{Superconductivity driven by pairing of the coherent parts of the physical electrons}

\author{Yuehua Su}
\email{suyh@ytu.edu.cn}
\affiliation{ Department of Physics, Yantai University, Yantai 264005, P. R. China}

\author{Chao Zhang}
\affiliation{ Department of Physics, Yantai University, Yantai 264005, P. R. China}

\begin{abstract}

How the superconductivity in unconventional superconductors emerges from the diverse mother 
normal states is still a big puzzle. Whatever the mother normal states are the superconductivity 
is {\em normal} with BCS-like behaviours of the paired quasiparticles in condensation. To reconcile 
the diverse mother normal states and the normal superconductivity in unconventional superconductors, 
we revisit a proposal that the emergence of the low-energy coherent parts of the physical electrons, 
which survive from the interaction correlations, is an essential prerequisite for superconductivity.  
The superconductivity is driven by the pair condensation of these coherent parts of the physical 
electrons. Moreover the incoherent parts of the physical electrons can enhance the superconducting 
transition temperature $T_c$ although they are not in driving role in the emergence of the 
superconductivity. Some experimental responses of the coherent parts of the physical electrons are
predicted. 

\end{abstract}



\maketitle

\section{Introduction}\label{sec1}

The generic phase diagram of the unconventional superconductors has a dome-structure superconducting 
phase. It is proximate to an antiferromagnetic insulator or bad-metal state on one underdoped side 
and to a normal Fermi-liquid state on the other overdoped side. The mother normal states from which 
the superconductivity emerges can be such as in the cuprate superconductors the mysterious pseudogap 
state, the strange metallic state or the normal Fermi-liquid state. 
Although the mother normal states are diverse with some in non-Fermi liquid states, it is remarkable 
that the emergent superconductivity is normal, where most of the superfluid responses such as the 
penetration depth, the Andreev reflection, the Josephson effects,  and the quasiparticle-relevant 
responses such as angle-resolved photoemission spectroscopy (ARPES), scanning tunneling microscope (STM), 
Raman scatterings, nuclear magnetic resonance (NMR), thermodynamics, impurity scattering, show BCS-like 
behaviours and can be simulated with BCS mean-field theoretical formalisms. 
This is confirmed recently by a universal scaling of the condensation energy in different-family 
superconductors~\cite{Stewart2015},
\begin{equation}
\frac{U}{\gamma}, \frac{U}{\Delta C /T_c} \propto  T_c ^{2} , \label{eq0}
\end{equation}
where $U$ is the condensation energy for superconductivity, $\gamma$ is the specific heat constant 
and $\Delta C$ is the specific heat jump at $T_c$. This universal scaling behaviour is followed in 
element superconductors, phonon-mediated metal superconductors, Fe-based superconductors, heavy fermion 
superconductors, Sr$_2$RuO$_4$, Li$_{0.1}$ZrNCl, organic superconductors and optimal-doped cuprate 
superconductors. This scaling behaviour shows that the superconductivity whatever the mother normal 
states are comes from the condensation of the Cooper pairs.  

How to reconcile the diverse mother normal states and the normal superconductivity is one big issue 
in the condensed matter filed. One interesting idea for this issue comes from Shen and Sawatzky, who 
propose that the superconducting phase transition in cuprate superconductors is not only the opening 
of the superconducting gap but also the emergence of the well-defined coherent quasiparticles~\cite{ShenPSS1999}. 
This is one idea in large debate, especially for underdoped cuprate superconductors which are mostly 
regarded in doped Mott-insulator states.
However, experiments seem to support this idea. The coherent weight of the physical electrons near 
antinodal points $\mathbf{k}=(\pi,0)/(0,\pi)$ accumulates largely when temperature decreases across 
$T_c$~\cite{FengShenScience2000,DingPRL2001}. The well-defined Fermi arc in the pseudogap normal state 
shows the emergence of the nodal quasiparticles~\cite{NormanCampuzano2006}, the coherent weight of which 
is largely suppressed with doping dependence much smaller than $2p/(p+1)$~\cite{FournierNatPhys2010}.   
The discrimination of the superconducting gap and the pseudogap as two distinct energy scales
~\cite{TanakaShenScience2006,LeeShenNature2007} strongly shows that the superconducting phase transition 
is relevant to the opening of the superconducting gap, not as coherence of the preformed Cooper pairs.  
Uchida's group find that the superconducting transition $T_c$ is smaller than both the coherent Fermi 
temperature $T_F$ and the pseudogap temperature $T^{\star}$ and conclude that the superconductivity 
emerges out of the coherent fermionic quasiparticles~\cite{UchidaPRB2009}.

In this article, we revisit this proposal that the emergence of the low-energy coherent parts of the 
physical electrons, which survive from the correlations of the electrons, is an essential prerequisite 
for superconductivity, and the superconductivity is driven by the condensation of the paired coherent 
parts of the physical electrons. We show that $T_c$ is strongly influenced by the renormalized coherent 
weight of the physical electrons and propose that this is one dominant mechanism for the doping variation 
of $T_c$ in underdoped unconventional superconductors. Our proposal is made for the unconventional 
superconductors, such as the cuprate superconductors, the Fe-based superconductors and the heavy fermion 
superconductors, etc. It should be noted that Anderson's resonating valence bond (RVB) theory~\cite{Anderson1987} 
and its extended gauge theories are one elegant formalism for cuprate superconductors. However, the novel 
spin-charge separation of the physical electrons in these theories is not confirmed by experiments, which 
instead favour the integrity of the physical electron as the component of the Cooper pair. 
There are no large debate on our proposal for the Fe-based superconductors where the interaction correlations 
are not strong. For the heavy fermion superconductors where antiferromagnetic quantum criticality may be 
relevant for superconductivity, the normal superconducting responses and the universal scaling behaviour 
in Eq. (\ref{eq0}) show that our proposal is one favour mechanism for superconductivity in the heavy 
fermion superconductors.

This article is arranged as following. In Sec. \ref{sec1}, we show a novel property in unconventional 
superconductors that the normal superconductivity can emerge from the diverse mother normal states, some 
of which are non-Fermi liquid states.  In Sec. \ref{sec2}, we revisit the proposal in detailed that the 
normal superconductivity in unconventional superconductors is driven by the condensed coherent parts of 
the physical electrons in pairs. In Subsection \ref{sec2.1}, we present a schematic formulation to 
redefine the mother normal states and the emergent superconducting state. In Subsection \ref{sec2.2}, 
we propose a single-particle spectrum function with the separated coherent and incoherent weights of 
the physical electrons in the mother normal states. In Subsection \ref{sec2.3}, we study the Thouless's 
instability for superconductivity and show how the coherent parts of the physical electrons drive the 
emergence of the superconductivity. The influence of the incoherent parts of the physical electrons on 
the superconductivity is also shown to enhance $T_c$ although they are not a driving factor.  
In Sec. \ref{sec3}, we present a reduced model for the renormalized coherent parts of the physical electrons, 
where we predict their responses in experiments. Sec. \ref{sec4} shows some remarks on the dominant factors 
on $T_c$. Some further theoretical problems are also discussed. In \ref{seca1}, some remarks are given on 
the theoretical formalisms, the weak-coupling BCS theory, the strong-coupling Eliashberg theory and the 
macroscopic Ginzburg-Landau functional theory.

\section{Superconductivity from mother normal states}\label{sec2}

\subsection{Schematic formulation for the mother normal states and the superconducting state}\label{sec2.1}

As there are no unified theory for the diverse mother normal states in unconventional superconductors, we 
introduce a schematic representation $\left\vert \Psi_m \right\rangle$ for them, 
\begin{equation} 
\left\vert \Psi_m \right\rangle = P_I \left\vert \Psi_0 \right\rangle . 
\label{eq1}
\end{equation}
Here $\left\vert \Psi_0 \right\rangle$ is the interaction-free ground state and $P_I$ is an operation to 
project the freely interacting ground state into a special mother normal state. $P_I$ is interaction relevant.

We should remark that the interactions in a model study are treated in different levels. The most relevant 
interactions in condensed matters are the electron-electron and electron-ion Coulomb interactions. While 
the static electron-ion Coulomb interaction has been included priorly in Bloch electronic band structure, 
the dynamical electron-phonon interactions should be involved exclusively. These interactions are the basic 
ones and in general have high energy scales. There are some other effective low-energy interactions such as 
super-exchange antiferromagnetic interactions which come from virtual processes. The relevant interactions 
in $P_I$ are the above two categories, which dominate the physics of the mother normal states.
The pairing interactions for superconductivity is separated from the above two categories in our formulation, 
although in concept they come from the projection of the above two-category interactions in particle-particle 
channels.

The mother normal states thus defined can be normal Fermi-liquid state, antiferromagnetic state with long-range 
order, metallic state with strong antiferromagnetic fluctuations, non-Fermi liquid state such as the strangle 
metallic state in the cuprate superconductors, the heavily renormalized fermionic state in the heavy fermion 
superconductors, etc. To establish well-defined theories or formalisms for the diverse novel mother normal 
states is still one challenge.  

The superconducting state from the mother normal states can be defined as 
\begin{equation}
\left\vert \Psi_{sc} \right\rangle = P_{sc} \left\vert \Psi_m \right\rangle , 
\label{eq2} 
\end{equation}
where $P_{sc}$ is the projection of the pairing interactions upon the mother normal state 
$\left\vert \Psi_m \right\rangle$ to produce the superconducting state $\left\vert \Psi_{sc} \right\rangle$.
The role of $P_{sc}$ is to bind the low-energy coherent parts of the physical electrons into pairs. In general 
case, the superconducting state $\left\vert \Psi_{sc} \right\rangle$ thus defined involves both the pairing 
physics of the low-energy coherent parts of the physical electrons and the normal physics of their high-energy 
parts. Superconducting phase transition is a pairing physics, which is our assumption and starting point to 
study the superconductivity in the unconventional superconductors.

In many cases, the pairing interactions in $P_{sc}$ are much smaller than the interactions in $P_I$. For 
example, the pairing attractive interaction in normal metal superconductors comes from the effective 
electron-phonon interactions. It has energy scale of the Debye frequency and is much smaller than the Fermi 
energy which comes from direct Coulomb interactions. The pairing interaction in heavy fermion superconductors 
is still in debate. The superconducting temperature $T_c \sim 1 K$, which is much smaller than the antiferromagnetic 
temperature $T_N\sim 100-1000 K$. The separation of the interaction energy scales in $P_{sc}$ for superconductivity
and in $P_{I}$ for mother normal states in these cases is the basic principle to preserve the reliability of 
our assumption in Eq.(\ref{eq2}). 
It should be noted that in our assumption $P_{sc} P_I \not= P_I P_{sc}$. Thus in our formulation the Anderson's 
RVB state defined by $\left\vert \Psi_{RVB} \right\rangle = P_I P_{sc} \left\vert \Psi_0 \right\rangle$ with 
$P_I$ the Hubbard-interaction driven Gutzwiller projection is not the physical relevant superconducting state 
in high-$T_c$ cuprate superconductors. Similarly, the scenario of the performed Cooper pairs in unconventional 
superconductors is also not relevant in our proposed formulation. Some remarks on the weak-coupling BCS theory 
and its strong-coupling version, the Eliashberg theory, are given in \ref{seca1}.

\subsection{Single-particle spectrum function of the mother normal states} \label{sec2.2}

As we have pointed out in the above sections, the superconductivity in unconventional superconductors is normal, 
i.e., the superconducting phase transition is a pairing physics and the superconducting responses in both 
superfluid and quasiparticle relevant channels show BCS-like behaviours. Moreover, the resistivity above $T_c$ 
involves the scattering effects of the interactions, which should not diminish absolutely when superconductors
enter into the superconducting state across $T_c$. It implies that the superconducting state involves both the 
low-energy dissipation-less superconducting condensate and the high-energy interaction scattering physics, 
most of the latter which are continuously transited from the mother normal states.
 
Based upon the above two facts, the normal superconductivity and the interaction scattering physics in 
superconducting state, we assume that in the mother normal states the single-particle spectrum function of 
the Green's function 
$G_{\sigma}(\mathbf{k},\tau) = -\langle T_{\tau} c_{\mathbf{k}\sigma}(\tau) c^{\dag}_{\mathbf{k}\sigma} (0) \rangle$
has two parts,
\begin{eqnarray}
A(\mathbf{k,\omega}) = A_{coh} (\mathbf{k,\omega}) + A_{inc} (\mathbf{k,\omega}) ,
\label{eq3}
\end{eqnarray}
where  $A_{coh} (\mathbf{k,\omega})$ is the low-energy coherent part defined by
\begin{eqnarray}
A_{coh}(\mathbf{k,\omega}) = Z_{\mathbf{k}} \delta_{\Gamma}
(\omega - \varepsilon^{\star}_{\mathbf{k}}) ,
\label{eq4}
\end{eqnarray}
and $A_{inc} (\mathbf{k,\omega})$ is the high-energy incoherent part. 
Here we have assumed that the normal state has no long-range order and the spectrum is spin independent.
$Z_{\mathbf{k}} \equiv Z_{\mathbf{k}} (\omega)|_{\omega = \varepsilon^{\star}_{\mathbf{k}}} $ is the coherent 
weight of the single physical electrons with renormalized energy $\varepsilon^{\star}_{\mathbf{k}}$.
$\delta_{\Gamma} (x) = \frac{1}{\pi} \frac{\Gamma_{\mathbf{k}}}{x^2 + \Gamma^{2}_{\mathbf{k}} }$
is an extended $\delta$-function with scattering rate $\Gamma_{\mathbf{k}}$.
Physically, the renormalized energy $\varepsilon^{\star}_{\mathbf{k}}$, the scattering rate $\Gamma_{\mathbf{k}}$ 
and the coherent weight $Z_{\mathbf{k}}$ of a physical electron are determined by the real scattering processes 
where the energy conservation law is preserved as described by the Fermi's Golden rule. The incoherent part 
$A_{inc}(\mathbf{k},\omega)$ comes from the virtual scattering processes where the energy conservation law 
is broken. In the latter case, the intermediate states for $c^{\dag}_{\mathbf{k}\sigma}$ during scattering 
may be a composite one such as defined by $\sum_{\mathbf{q}} c^{\dag}_{\mathbf{k-q},\sigma} b^{\dag}_{\mathbf{q}}$ 
with $b^{\dag}_{\mathbf{q}} = \sum_{\mathbf{k_1}\sigma} c^{\dag}_{\mathbf{k_1+q},\sigma} 
c_{\mathbf{k_1}\sigma} + \cdots $. 

The coherent parts of the physical electrons have dominant roles in the normal pairing superconductivity.
The renormalized electronic structure $\varepsilon^{\star}_{\mathbf{k}}$ has a well-defined Fermi surface, 
whose topology combined with the pairing interaction plays important role in the symmetry of the pairing gap 
potential~\cite{HuJ}. The finite coherent weight $Z_{\mathbf{k}}$ has a dominant role in determination of 
$T_c$ as shown below. 

The incoherent part $A_{inc}(\mathbf{k,\omega})$ describes the high-energy relevant physics in the single-particle 
channels and has negligible modification in superconducting phase transition from the mother normal states to 
the superconducting state. It should be noted that the physical electrons in the mother normal states are
strongly entangled even in the normal Fermi-liquid state. The entanglements of the physical electrons which 
contribute to the incoherent part $A_{inc}(\mathbf{k,\omega})$ is an exact $N$-body problem unsolved.

\subsection{Thouless's criterion for superconductivity} \label{sec2.3}

In this article, we propose that the normal superconductivity is driven by the pairing of the coherent 
parts of the physical electrons and the finite coherent weight $Z_{\mathbf{k}}$ near Fermi energy is 
essential prerequisite for superconductivity in unconventional superconductors. 

We will use the Thouless's criterion to study the superconductivity instability of the mother normal 
states~\cite{Thouless1960,Nozieres1985}. Consider a simple case where the attractive pairing interaction 
is dominantly in channel with symmetry function $\phi_{\mathbf{k}} $, 
\begin{eqnarray}
H_{p} = -\frac{1}{N} \sum_{\mathbf{k_1 k_2}} g \phi_{\mathbf{k}_1} \phi_{\mathbf{k}_2} 
c^{\dag}_{\mathbf{k_1}\uparrow} c^{\dag}_{-\mathbf{k_1}\downarrow}
c_{-\mathbf{k_2}\downarrow} c_{\mathbf{k_2}\uparrow} . \label{eq5}
\end{eqnarray}
Introduce the pairing operator 
$$\Delta = \frac{1}{\sqrt{N}} \sum_{\mathbf{k}} \phi(\mathbf{k}) 
c_{-\mathbf{k}\downarrow} c_{\mathbf{k}\uparrow} ,  
$$ 
we define the pairing susceptibility as
\begin{equation}
\chi (\tau) = \langle T_{\tau} \Delta(\tau) \Delta^{\dag}(0) \rangle .
\label{eq6}
\end{equation}
In random-phase approximation (RPA), the pairing susceptibility is given by
\begin{equation}
\chi (i\nu_n) = \frac{\chi_0 (i\nu_n)} {1 - g \chi_0 (i\nu_n)} ,
\label{eq7}
\end{equation}
where $\chi_0 (i\nu_n)$ is the bare zero-th order pairing susceptibility. The Thouless's criterion for 
superconductivity is defined as~\cite{Thouless1960,Nozieres1985}
\begin{equation}
1-g \chi_0 ( 0 ) = 0 . \label{eq8}
\end{equation}

Suppose the spectrum function of the mother normal states 
$A(\mathbf{k},\omega) = A_{coh} (\mathbf{k},\omega) + A_{inc} (\mathbf{k},\omega)$ has following behaviours,
\begin{equation}
A_{coh} (\mathbf{k},\omega) = Z_{\mathbf{k}}
\delta(\omega - \varepsilon^{\star}_{\mathbf{k}}),
 A_{inc} (\mathbf{k},\omega) = P_{\mathbf{k}} \theta_{\Lambda} (\omega) ,
 \label{eq9}
\end{equation}
where $\theta_{\Lambda} (\omega) =\frac{1}{2\Lambda} \left[\theta(\omega+\Lambda) - \theta(\omega-\Lambda)\right]$
with $\theta(x)$ the step function and $\Lambda$ the effective bandwidth, and $P_{\mathbf{k}} = 1 - Z_{\mathbf{k}}$.
It can be shown that
\begin{eqnarray}
\chi_0 (0) = \chi_0^{(1)} + \chi_0^{(2)} +  \chi_0^{(3)} , \label{eq10}
\end{eqnarray}
with
\begin{eqnarray}
\chi_0^{(1)} &=& \frac{1}{N}\sum_{\mathbf{k}} \left(\phi_{\mathbf{k}} Z_{\mathbf{k}}\right)^{2}
\frac{f_{\beta}(\varepsilon^{\star}_{\mathbf{k}})}{\varepsilon^{\star}_{\mathbf{k}}} , \nonumber \\
\chi_0^{(2)} &=& \frac{2}{N}\sum_{\mathbf{k}} \phi^{2}_{\mathbf{k}} Z_{\mathbf{k}} P_{\mathbf{k}}
\int_{-\infty}^{+\infty} d\omega \theta_{\Lambda}(\omega) 
\frac{f_{\beta}(\varepsilon^{\star}_{\mathbf{k}})+f_{\beta}(\omega)}{\varepsilon^{\star}_{\mathbf{k}}+\omega}
 , \nonumber \\
\chi_0^{(3)} &=& \frac{1}{N}\sum_{\mathbf{k}} \left(\phi_{\mathbf{k}} P_{\mathbf{k}}\right)^{2}
\int_{-\infty}^{+\infty} d\omega_1 d\omega_2 \theta_{\Lambda}(\omega_1) \theta_{\Lambda}(\omega_2) \nonumber \\
&& \hspace{0.5cm} \cdot \frac{f_{\beta}(\omega_1)+f_{\beta}(\omega_2)}{\omega_1+\omega_2} . \nonumber
\end{eqnarray}
Here $f_{\beta}(\omega) \equiv  \frac{\tanh\left( \beta\omega/2 \right)} {2}$.
Physically, $\chi_0^{(1)}$ comes from the pure coherent parts of the physical electrons, $\chi_0^{(2)}$ 
is the mixed contribution of the coherent and incoherent parts and $\chi_0^{(3)}$ is the contribution from 
the pure incoherent parts. One additional interesting result is that the renormalized factors $Z_{\mathbf{k}}$ 
and $P_{\mathbf{k}}$ appear combined with the pairing symmetry function $\phi_{\mathbf{k}}$. It implies 
that the pairing symmetry function, which is dominated by the pairing interaction, can be renormalized 
further by the renormalized factors. This is an interesting result analogue to the strong-coupling Eliashberg 
theory, as remarked in Sec. \ref{sec3}.
It should be noted that in the above Thouless's approximation for the superconducting instability, we only 
consider the most relevant ladder-diagram RPA contribution in the pairing susceptibility. Moreover, we have 
defined approximately the single-particle spectrum function following Eq. (\ref{eq9}), which involves the 
universal feature of one finite coherent peak and one broad incoherent background as observed in ARPES. 
All other novel physics of the normal states are assumed irrelevant to the superconducting instability and 
are ignored in our study.   

Consider a simple s-wave case with $\mathbf{k}$-independent renormalized factor, 
\begin{equation}
\phi_{\mathbf{k}} = 1, Z_{\mathbf{k}} = Z, \label{eq12}
\end{equation}
and in a weak-coupling limit where the pairing interaction $g$ is mainly within an energy range 
$\omega_g \ll \Lambda$. In this case, $\chi_0^{(1)}$ shows singularity at low temperature $T\rightarrow 0^{+}$,
\begin{equation}
\chi_0^{(1)} = Z^{2} N_F \ln \left(\frac{\omega_g}{0^{+}}\right) . \label{eq13}
\end{equation}
However, in the same limit, $\chi_0^{(2)}$ and $\chi_0^{(3)}$ shows non singularity,
\begin{eqnarray}
\chi_0^{(2)} & = & 4(\ln 2) Z P N_F\frac{\omega_g}{\Lambda} , \label{eq14}  \\
\chi_0^{(3)} & = & (\ln 2) P^2 \frac{\omega_g}{\Lambda^2} . \label{eq15}
\end{eqnarray}
Here $N_F$ is the density of state (DOS) at Fermi energy of the mother normal state with renormalized dispersion 
$\varepsilon_{k}^{\star}$ and $P=1-Z$. The logarithm singularity of $\chi_0^{(1)}$ implies that at some finite 
$T_c$, the Thouless's condition of the superconductivity Eq. (\ref{eq8}) can be satisfied. $T_c$ is the critical 
temperature for superconducting phase transition. Since there are non singularity in $\chi_0^{(2)}$ and 
$\chi_0^{(3)}$, the superconductivity is dominantly driven by the coherent parts of the physical electrons. 
As $\chi_0^{(2)}$ and $\chi_0^{(3)}$ are shown to be positive, the Thouless's criterion Eq. (\ref{eq8}) 
can be modified into the form:
\begin{equation}
\langle 1\rangle_{T} -g \chi_0^{(1)} = 0 , \label{eq16}
\end{equation}
where $\langle 1\rangle_T$ is a renormalized value by the incoherent parts of the physical electrons defined 
by $\langle 1\rangle_T \equiv 1 - g (\chi_0^{(2)} + \chi_0^{(3)})$. The superconducting transition $T_c$ follows
\begin{equation}
T_c = \alpha \omega_g e^{-\frac{\langle 1\rangle_{T_c}}{g Z^{2} N_F}}, \label{eq17}
\end{equation} 
where the constant $\alpha \simeq 1.13  $ for $s$-wave attractive interaction.

There are two dominant factors for $T_c$, one comes from the pairing interaction, such as the interaction 
range $\omega_g$, the interaction constant $g$ and the pairing symmetry $\phi_{\mathbf{k}}$,  and the other 
comes from the qusiparticle properties of the mother normal state, such as the renormalized DOS and the 
coherent weight $Z_{\mathbf{k}}$. Larger pairing interaction range and stronger interaction constant favour 
higher $T_c$. More normal DOS and larger coherent weight lead to higher $T_c$. One additional interesting 
factor comes from the incoherent parts of the physical electrons in $\langle 1\rangle_T$. Although it is 
not a driving factor for superconductivity, the renormalization effects of the incoherent parts of the 
physical electrons can enhance $T_c$.

In the above weak-coupling limit, the renormalized DOS can be approximated constant near Fermi energy. 
In the reverse strong-coupling limit with $\omega_g \gg \Lambda$, if the DOS is weakly energy-dependent 
in such as quasi-two-dimensional system, the pairing susceptibilities $\chi_0^{(i)}$ have similar singularity 
behaviours with just a substitution of $\omega_g$ by $\Lambda$. In this case, the coherent parts of the 
physical electrons can also lead to the dominant singularity of the superconducting pair instability. A 
general strong-coupling limit with strong frequency-dependent DOS should be studied extensively\cite{Gorkov2015} 
as there maybe a crossover from wide-band BCS to narrow-band BEC~\cite{Nozieres1985,MicnasRMP1990}.

The mother normal states of the underdoped unconventional superconductors have more abnormal physics beyond 
the Fermi-liquid state. The interaction correlations strongly suppress the coherent parts of the physical 
electrons and thus reduce the coherent weight $Z_{\mathbf{k}}$. The renormalized reduced $Z_{\mathbf{k}}$ 
largely suppresses the superconducting temperature $T_c$. With doping increases, the electron correlations 
become weaker and $Z_{\mathbf{k}}$ increases. This leads to an enhanced $T_c$ with increased doping. 
Therefore, the doping variation of $T_c$ in underdoped unconventional superconductors is strongly influenced 
by the interaction renormalization coherent weight $Z_{\mathbf{k}}$.
   
\section{A reduced model for the renormalized normal superconductivity}\label{sec3}

In this section, we consider a simple case where the roles of $P_I$ can be approximated by a coherent projection. 
Consider a superconductor with Hamiltonian $H=H_t + H_p$, where the kinetic part $H_t$ is given by 
$H_t = \sum_{\mathbf{k}\sigma}\varepsilon_{\mathbf{k}} c^{\dag}_{\mathbf{k}\sigma} c_{\mathbf{k}\sigma}$
and the pairing interaction $H_p$ is defined by Eq. (\ref{eq5}). Here the interactions in $P_I$ of Eq. (\ref{eq1}) 
are not explicitly given out. The renormalization effects of the projection $P_I$ on the Hamiltonian $H$ can 
be described by $\widetilde{H} = P_I^{\dag} H P_I$. Consider the following projection of the operator,
\begin{equation}
P_I^{\dag} c_{\mathbf{k}\sigma} P_I = Z_{\mathbf{k}}^{1/2}  c_{\mathbf{k}\sigma} + \cdots , 
\label{eq18} 
\end{equation}
the coherent part of the renormalized Hamiltonian $\widetilde{H}$ can be simplified as 
\begin{equation}
\widetilde{H}_{coh} = \widetilde{H}_{t} + \widetilde{H}_{p}, \label{eq19} 
\end{equation}
where  $\widetilde{H}_{t} $ and $\widetilde{H}_{p} $ are given by
\begin{eqnarray}
\widetilde{H}_{t} &=& \sum_{\mathbf{k}\sigma} \varepsilon^{\star}_{\mathbf{k}} 
c^{\dag}_{\mathbf{k}\sigma} c_{\mathbf{k}\sigma} , \label{eq20}\\
\widetilde{H}_{p} &=& -\frac{1}{N} \sum_{\mathbf{k_1 k_2}} \widetilde{g} 
\widetilde{\phi}_{\mathbf{k}_1} \widetilde{\phi}_{\mathbf{k}_2} 
c^{\dag}_{\mathbf{k_1}\uparrow} c^{\dag}_{-\mathbf{k_1}\downarrow}
c_{-\mathbf{k_2}\downarrow} c_{\mathbf{k_2}\uparrow} . \nonumber
\end{eqnarray}
The renormalized variables are defined as following: 
\begin{equation}
\varepsilon^{\star}_{\mathbf{k}} = Z_{\mathbf{k}} P_I^{\dag}\varepsilon_{\mathbf{k}} P_I, 
\widetilde{g} = P_I^{\dag} g P_I ,
\widetilde{\phi}_{\mathbf{k}} = Z_{\mathbf{k}}P_I^{\dag}\phi_{\mathbf{k}}P_I . \label{eq21}
\end{equation}
It should be noted that there are two renormalization factors in the band dispersion $\varepsilon^{\star}_{\mathbf{k}}$,
one comes from the projection coherent weight $Z_{\mathbf{k}}$ and the other, denoted as $P_I^{\dag}\varepsilon_{\mathbf{k}} P_I$, 
stems from other interaction effects of $P_I$. These two factors are also relevant to the renormalization of DOS. 

The above coherent projection approximation comes from the following assumption for the ground state: 
\begin{equation}
\langle \Psi_{sc}\vert H \vert \Psi_{sc}\rangle 
\simeq \langle \Psi_{0}\vert  \widetilde{H}_{coh}  \vert \Psi_{0}\rangle
\simeq \langle \Psi^{(0)}_{sc}\vert  \widetilde{H}_{coh}  \vert \Psi^{(0)}_{sc}\rangle . \label{eq22}
\end{equation}
Here $\vert \Psi^{(0)}_{sc}\rangle$ is the weak-coupling BCS mean-field ground state corresponding to 
the renormalized coherent Hamiltonian $\widetilde{H}_{coh}$. In the above derivation, we use 
$P^{\dag}_{sc} H P_{sc} = H$ since only the pairing interaction $H_p$ involved in Hamiltonian $H$. 

Now let us extend the above approximation at ground state into the finite-temperature case. For any operator 
$O$, the thermal average $\langle O \rangle$ is defined and approximated as following:
\begin{eqnarray}
\langle O \rangle &\equiv & \frac{T_r \left[ e^{-\beta (H+H_I) } O \right]}
{T_r \left[e^{-\beta (H+H_I)}\right]}  \nonumber \\
 & \simeq & \frac{T_r \left[P^{\dag}_I e^{-\beta H } O P_I \right] }
{T_r \left[ P^{\dag}_I  e^{-\beta H} P_I\right] }  \nonumber \\
 & \simeq & \langle \widetilde{O}\rangle_{coh} , \label{eq23}
\end{eqnarray}
where $\widetilde{O} = P^{\dag}_I O P_I$ and 
$\langle \widetilde{O}\rangle_{coh} \equiv T_r\left[\widetilde{\rho}_{coh} \widetilde{O} \right]$
with $\widetilde{\rho}_{coh} = \frac{e^{-\beta \widetilde{H}_{coh} }} 
{T_r \left[e^{-\beta \widetilde{H}_{coh}}\right]} $. In the above approximation, the interactions in $P_I$ 
are only approximated by the projection operation $P_I$ and the high-energy incoherent physics is neglected.
In our following discussion, the coherent approximation in Eq. (\ref{eq22}) and Eq. (\ref{eq23}) is assumed. 

\subsection{Pairing order parameter}

Introduce a pairing order parameter $\Delta_c$ as
\begin{equation}
\Delta_c = - \frac{1}{N} \sum_{\mathbf{k}} \widetilde{g} \widetilde{\phi}_{\mathbf{k}} 
\langle c_{-\mathbf{k}\downarrow} c_{\mathbf{k}\uparrow} \rangle , \label{eq24}
\end{equation}
we can decouple the renormalized pairing interaction $\widetilde{H}_{p}$ into a mean-field approximation 
\begin{equation}
\widetilde{H}_{p} = \sum_{\mathbf{k}} \Delta_{\mathbf{k}}
\left( c^{\dag}_{\mathbf{k}\uparrow} c^{\dag}_{-\mathbf{k}\downarrow} + 
c_{-\mathbf{k}\downarrow} c_{\mathbf{k}\uparrow} \right) , \label{eq25}
\end{equation}
where  $\Delta_{\mathbf{k}} = \Delta_c \widetilde{\phi}_{\mathbf{k}}$. We can then obtain a self-consistent 
equation for the pairing order parameter, 
\begin{equation}
1 = \frac{1}{N} \sum_{\mathbf{k}} \frac{\widetilde{g} \widetilde{\phi}_{\mathbf{k}}^{2}}{2 E_{\mathbf{k}}}
\tanh \left(\frac{\beta E_{\mathbf{k}}}{2}\right) , \label{eq26}
\end{equation}
where $E_{\mathbf{k}} = \sqrt{\varepsilon^{* 2}_{\mathbf{k}}+\vert\Delta_{\mathbf{k}}\vert^{2}}$ is the 
energy of the coherent parts of the Bogoliubov quasiparticles. 

Eq. (\ref{eq26}) shows that the pairing order parameter $\Delta_c$ has a BCS mean-field solution. Moreover, 
the pairing symmetry function $\Delta_{\mathbf{k}}$ is renormalized by the coherent weight $Z_{\mathbf{k}}$.
This latter fact is important since it shows that the pairing symmetry function $\Delta_{\mathbf{k}}$ observed 
in ARPES is not only determined by the pairing interaction but also renormalized by the coherent weight. 
This result is analogue to that in the strong-coupling Eliashberg theory. If the higher-order renormalization 
effects of the interactions in $P_I$ can be ignored, it predicts from Eq. (\ref{eq26}) that
\begin{equation}
\frac{\Delta_m}{T_c} = const. \label{eq27} 
\end{equation}
where $\Delta_m \equiv \Delta_c \vert\widetilde{\phi}_{\mathbf{k}}\vert_{max}$ and $T_c$ is given by 
Eq. (\ref{eq17}) with $\langle 1\rangle_{T_c} \rightarrow 1$.

In cuprate superconductors, ARPES shows that the pairing symmetry function has exact d-wave symmetry~\cite{LeeShenNature2007,ZhouXJPRB2009}, $\Delta_{\mathbf{k}} \sim \left(\cos k_x - cos k_y\right)$. 
It is consistent exactly to the theoretical prediction from the hypothesis that the nearest-neighbour 
antiferromagnetic interaction is the pairing interaction. This result shows that the coherent weight 
$Z_{\mathbf{k}}$ in cuprate superconductors is {\em unusual} $\mathbf{k}$ independent, 
\begin{equation}
Z_{\mathbf{k}} = Z .  \label{eq28}
\end{equation}
The $\mathbf{k}$-independent coherent weight $Z$ implies that the dominant interactions in $P_I$ are on-site 
ones, which seems rule out the antiferromagnetic fluctuations with characteristic momentum $\mathbf{Q}=(\pi,\pi)$ 
as the dominant interactions for the abnormal mother normal states in cuprate superconductors. 

\subsection{Experimental responses}

Firstly, let us consider the superfluid responses in the penetration depth measurement. The diamagnetic 
current corresponding to the model Hamiltonian $H$ is given by
$j^{(d)}_{\alpha} (\mathbf{q}) = - \Lambda_{\alpha} A_{\alpha}(\mathbf{q})$,
where $\mathbf{A}(\mathbf{q})$ is the vector potential for electromagnetic filed, and 
$\Lambda_{\alpha} = \sum_{\mathbf{k}\sigma} T_{\alpha}(\mathbf{k})
\langle c^{\dag}_{\mathbf{k}\sigma} c_{\mathbf{k}\sigma} \rangle $ with
$T_{\alpha}(\mathbf{k})$ a kinetic energy $\varepsilon_{\mathbf{k}}$ relevant parameter. It can be shown 
that the penetration depth $\lambda_{\alpha}$ follows
\begin{equation}
\lambda_{\alpha} =\frac{1}{\sqrt{\mu_0 \widetilde{\Lambda}_{\alpha}} } , \label{eq29}
\end{equation}
where $\widetilde{\Lambda}_{\alpha} = \sum_{\mathbf{k}\sigma} \widetilde{T}_{\alpha}(\mathbf{k})
Z_{\mathbf{k}}\langle c^{\dag}_{\mathbf{k}\sigma} c_{\mathbf{k}\sigma} \rangle_{coh}$ 
with $\widetilde{T}_{\alpha}(\mathbf{k}) = P^{\dag}_I T_{\alpha}(\mathbf{k}) P_I$, and $\mu_0$ is the vacuum 
permeability. For the case with  $Z_{\mathbf{k}}$ is nearly $\mathbf{k}$ independent, we have 
\begin{equation}
\lambda_{\alpha} = \frac{\lambda_{\alpha,coh}}{\sqrt{Z}}, \label{eq30}
\end{equation}
where $\lambda_{\alpha,coh} = \frac{1}{\sqrt{\mu_0 \widetilde{\Lambda}_{\alpha, coh}} } $
with $\widetilde{\Lambda}_{\alpha, coh} = \sum_{\mathbf{k}\sigma} \widetilde{T}_{\alpha}(\mathbf{k})
\langle c^{\dag}_{\mathbf{k}\sigma} c_{\mathbf{k}\sigma} \rangle_{coh}$. 

In cuprate superconductors, since the nodal Fermi velocity is nearly doping independent~\cite{LanzaraShenNature2001,ZhouShenNature2003}, the renormalized low-energy dispersion and the renormalized 
coefficient of the diamagnetic current $Z_{k} \widetilde{T}_{\alpha}(\mathbf{k})$ should be both nearly doping independent. 
Therefore, in cuprate superconductors, it seems that the doping dependence of $\lambda_{\alpha}^{-2}$ should only 
be dominated by the charge-carrier density and the renormalization influence from $P_I$ could be nearly neglectable.
Why the linear doping dependence of the coherent weight $Z$ can coexist with the weakly linear doping dependence
of the nodal Fermi velocity in underdoped cuparates is still one mysterious problem. A formal study on this issue 
is one future subject. 

Secondly, let us consider the low-energy and low-temperature thermodynamic responses. The free-energy 
corresponding to the coherent Hamiltonian $\widetilde{H}_{coh}$ is given by 
$\widetilde{F}_{coh} = -\frac{1}{\beta} \ln T_r \left[e^{-\beta \widetilde{H}_{coh}} \right]$.
In the superconducting state with finite pairing order parameter, the free-energy is approximated further as 
\begin{equation}
\widetilde{F}^{(0)}_{coh} = -\frac{1}{\beta} \ln T_r \left[e^{-\beta \widetilde{H}_{mf}} \right] ,
\label{eq31}
\end{equation}
where $\widetilde{H}_{mf}$ is the corresponding superconducting mean-field Hamiltonian of $\widetilde{H}_{coh}$. 
Since all of the thermodynamic variables such as the entropy $S$ and the specific heat $C$ etc. can be 
obtained by the derivatives of the free energy, the renormalization effects of the interactions in $P_I$ 
on the thermodynamic responses are dominantly manifested by the energy spectrum $E_{\mathbf{k}}$ and there 
are no additional renormalized factor such as $\frac{1}{\sqrt{Z}}$ in penetration depth of Eq. (\ref{eq30}). 

Thirdly, let us consider the single-particle responses in such as ARPES and STM experiments. For the 
single-particle Green's function $G_{\sigma}(\mathbf{k},\tau) = -\langle T_{\tau} c_{\mathbf{k}\sigma}(\tau)
c^{\dag}_{\mathbf{k}\sigma} (0) \rangle$, in the coherent approximation defined by Eq. (\ref{eq22}) and 
Eq. (\ref{eq23}), we have 
\begin{equation}
G_{\sigma}(\mathbf{k},\tau) = Z_{\mathbf{k}} \widetilde{G}_{\sigma,coh}(\mathbf{k},\tau) 
\label{eq32}
\end{equation}
where $\widetilde{G}_{\sigma,coh}(\mathbf{k},\tau) = -\langle T_{\tau} c_{\mathbf{k}\sigma}(\tau)
c^{\dag}_{\mathbf{k}\sigma} (0) \rangle_{coh}$. Therefore, the spectrum function in ARPES 
$A(\mathbf{k},\omega)=-\frac{1}{\pi} \Im G_{\sigma}(\mathbf{k},\omega)$ is approximated at low-energy 
in superconducting state as
\begin{equation}
A(\mathbf{k},\omega) = Z_{\mathbf{k}} \widetilde{A}_{coh} (\mathbf{k},\omega), \label{eq33}
\end{equation}
where $\widetilde{A}_{coh}(\mathbf{k},\omega) = u_{\mathbf{k}}^{2}\delta_{\Gamma}(\omega-E_{\mathbf{k}})
+ v_{\mathbf{k}}^{2}\delta_{\Gamma}(\omega+E_{\mathbf{k}} ) $ and $u_{\mathbf{k}}^{2}, v_{\mathbf{k}}^{2}
 = \frac{1}{2} \left(1\pm \frac{\varepsilon^{\star}_{\mathbf{k}}}{E_{\mathbf{k}}}\right)$. 

The STM experimental signal is related to the local DOS $\rho(\omega) = \frac{2}{N} \sum_{\mathbf{k}} A(\mathbf{k},\omega) $.
In the coherent approximation, the low-energy local DOS is approximate as 
\begin{equation}
\rho(\omega) = Z \widetilde{\rho}_{coh} (\omega) , \label{eq34} 
\end{equation}
where $\widetilde{\rho}_{coh} (\omega) =\frac{2}{N} \sum_{\mathbf{k}} \widetilde{A}_{coh} (\mathbf{k},\omega) $. 
In Eq. (\ref{eq34}) we have considered a simple case that the coherence weight $Z_{\mathbf{k}}$ is $\mathbf{k}$ 
independent. 

Lastly, let us focus on the low-energy and low-temperature magnetic responses in such as Knight shift and 
$1/T_1 T$ from NMR and the inelastic spectra from neutron scatterings. These magnetic responses are related 
to the spin susceptibility defined by 
$\chi^{(s)}_{\alpha\beta} (\mathbf{q},\tau) = \langle T_{\tau} S_{\alpha}(\mathbf{q},\tau) 
S_{\beta}(-\mathbf{q},0)\rangle$, where the spin operator $S_{\alpha}(\mathbf{q})
=\frac{1}{\sqrt{N}}\sum_{\mathbf{k}\sigma\sigma^{\prime}}
c^{\dag}_{\mathbf{k}\sigma}\left(\frac{\boldsymbol{\sigma}}{2}\right)_{\sigma\sigma^{\prime}}
c_{\mathbf{k}+\mathbf{q}\sigma^{\prime}}$. It is easily shown that in the coherent approximation and with 
$\mathbf{k}$-independent quasiparticle coherent weight, the spin response susceptibility at low-energy and 
low-temperature is approximate as
\begin{equation}
\chi^{(s)}_{\alpha\beta} (\mathbf{q},\tau) = Z^{2} 
\widetilde{\chi}^{(s)}_{\alpha\beta, coh} (\mathbf{q},\tau) , \label{eq35}
\end{equation} 
where $\widetilde{\chi}^{(s)}_{\alpha\beta, coh} (\mathbf{q},\tau)= \langle T_{\tau} S_{\alpha}(\mathbf{q},\tau) 
S_{\beta}(-\mathbf{q},0)\rangle_{coh}$. It should be noted that only the contribution from the itinerant coherent 
parts of the physical electrons has been considered in Eq. (\ref{eq35}) to the low-energy and low-temperature 
magnetic responses, which will be largely suppressed in superconducting state due to the opening of the 
superconducting gap.   

\subsection{Remarks}

In the above discussion, the coherent parts of the physical electrons are simply described by a reduced coherent 
Hamiltonian $\widetilde{H}_{coh}$ defined in Eq.(\ref{eq19}), where the higher-order renormalization effects 
other than the projection $P_I$ from the incoherent parts of the physical electrons are neglected. For realistic 
superconductors, when the physical electrons are in strong dynamical entanglement, the roles of the interactions 
in $P_I$ may not be well described by the simple coherent projection approximation defined in Eq. (\ref{eq22}) 
and Eq. (\ref{eq23}). In this case, the entanglement physics requires a theory where both the normal 
superconductivity and its strongly correlated mother normal states should be involved, such as the strong-coupling 
Eliashberg theory for the superconductors with a Fermi-liquid mother normal state. 
 
Before the end of this section, some remarks on the renormalization in the RVB formalisms may be useful. In the 
RVB formalisms, the strong Mott-Hubbard interaction correlations are assumed to the driven factors for both the 
unconventional superconductivity and the abnormal mother normal states in cuprate superconductors~\cite{Anderson1987,AndersonRVB2004,PALeeRMP2006}.
In a simplified treatment on the t-J model, two renormalized factors are introduced to treat the Mott-Hubbard
Gutzwiller projection effects~\cite{ZhangRice1988}. One is $g_t = \frac{2p}{1+p}$ for the kinetic energy and the 
other is $g_s = \frac{4}{(1+p)^2}$ for the spin-spin Heisenberg energy ($p$ is the hole doping concentration). 
$g_t$ introduces a new energy scale which renormalizes the superconducting pairing gap function with linear 
doping dependence in underdoped cuprates. In a more formal slave-boson theory for the RVB, the physical electron 
is assumed to be decoupled into charge-carrier holon and spin-carrier spinon~\cite{PALeeRMP2006}. Superconductivity 
can emerge when the spinons are paired and the holons are in coherent motion with Bose-Einstein condensation. 
Thus there are two different mechanisms for the superconducting phase transition in cuprates.
While the superconducting phase transition in overdoped case is driven by the pairing formation of the spinons, 
it is dominated by the coherent condensation of the holons in the underdoped case. In the latter case, the pairing 
gap and the superconducting phase transition $T_c$ are two independent energy scales. This is basically different 
to our ideas in this article, here we assume that in all cases, the superconducting $T_c$ is one energy scale for 
the emergence of the pairing of the coherent parts of the physical electrons. It should be noted that the brilliant 
RVB proposal and the various relevant theoretical formalisms are still in doubt, as the predicted spinons and holons 
have not been found in experiments.

\section{Discussions and remarks} \label{sec4}

In this article, we propose that the finite coherent parts of the physical electrons in the mother normal states 
are prerequisite for superconductivity. By considering the Thouless's criterion for superconductivity instability, 
we show that the finite coherent parts of the physical electrons drive the emergence of the superconductivity. 
The superconducting $T_c$ is determined by two factors, the attractive pairing interaction and the quasiparticle 
properties of the physical electrons. The former involves the interaction range, the interaction constant and the 
pairing symmetry function; the latter involves the coherent weight $Z_{\mathbf{k}}$, the DOS and the Fermi-surface 
topology of the electronic structure $\varepsilon^{\star}_\mathbf{k}$. Larger interaction range and larger interaction 
constant favour higher $T_c$, and more renormalized DOS and larger $Z_{\mathbf{k}}$ lead to higher $T_c$. 
Moreover, although the incoherent parts of the physical electrons play no dominant role for superconductivity, 
they can also enhance $T_c$ by a renormalization effect. We show that the reduced coherent weight $Z_{\mathbf{k}}$ 
can be one dominant factor for the superconducting phase transition in underdoped unconventional superconductors. 

It should be noted that the collective modes of the superconducting condensate may be another important factor 
for superconductivity. Most of the unconventional superconductors are quasi-two-dimensional in crossover from 
two- to three-dimensions. Hohenberg theory~\cite{Hohenberg1967} states that there is no finite-$T_c$ superconductivity 
in one- and two-dimensional superconductors since the collective modes can easily destroy the long-range 
superconducting order. The roles of the collective modes of the superconducting condensate have been investigated by 
Emery and Kivelson as a proposal to account for the reduction of $T_c$ in underdoped cuprate superconductors~\cite{EmeryKivelson}. 
Which one, the reduced coherent weight of the physical electrons or the thermal fluctuations of the collective modes, 
plays the dominant roles in the reduction of $T_c$ in unconventional superconductors is still an issue to be studied.

It is known that the well-established BCS formalisms, the weak-coupling BCS theory and the strong-coupling Eliashberg 
theory, can not predict reliable $T_c$ for a given superconductor. From our study we can understand why the prediction 
of $T_c$ is so unreliable. Most of the attractive pairing interactions for the Cooper pairs are residual ones generated 
from high-energy scattering processes and thus are hardly defined definitely for a realistic superconductor. The 
single-particle coherent weight $Z_{\mathbf{k}}$ and the renormalized band structure $\varepsilon^{\star}_\mathbf{k}$
are strongly correlated to the high-energy physics in the mother normal states where a well-defined theory is absent. 
The feedback effects of the collective modes of the paired condensate involve  dimensionality. All these factors in a 
realistic superconductor are too complex to be definitely defined.

In future, we have some further problems to be considered following our proposal, such as to study the experimental 
responses of the paired coherent parts of the physical electrons which drive the emergence of the superconductivity 
and to show how these coherent parts survive from the correlated diverse mother normal states.  

It should also be noted that in this article we introduce the single-particle spectrum function $A(\mathbf{k},\omega)$ 
to describe the low-energy coherent physics and the high-energy incoherent entanglement of the physical electrons. 
One underlying well-established formalism is the strong-coupling Eliashberg theory where the extended self-energy and 
the full single-particle spectrum function of the Nambu's spinor for the mixed particles and holes are self-consistently 
constructed. However in the Eliashberg theory the mother normal state is assumed the Fermi-liquid state. Moreover 
only the single-particle excitations are involved directly in the Eliashberg theory and the underlying ground state 
of the superconductors are not known as discussed in \ref{seca1}. More exact definition and better well-defined 
formalisms to describe the coherent and incoherent physics of the physical electrons are worth to be established.

\section{Acknowledgements}
This work was supported by the National Natural Science Foundation of China
(Grant Nos. 11304269, 11304268 and 11774299) and the Natural Science Foundation
of Shandong Province (Grant No. ZR2017MA033).

\appendix 

\section{Remarks on weak-coupling BCS theory, strong-coupling Eliashberg theory
and Ginzburg-Landau functional theory} \label{seca1}

There are some well-established theories for superconductivity, the weak-coupling BCS theory and the 
strong-coupling Eliashberg theory in BCS formalism from microscopic perspective, and the Ginzburg-Landau 
functional theory from the macroscopic perspective. In the weak-coupling BCS theory, the mother normal state 
is approximated as Fermi-liquid state. The $P_I$ is now defined as the perturbation evolution $S$-matrix, 
$P_I = S(0,-\infty)$. The Fermi-liquid ground state is adiabatically evolved from the interaction-free ground 
state and the low-energy excitations are the so-called {\em quasiparticles}, which can be regarded as freely 
independent excitations with interaction renormalization included. The superconductivity in weak-coupling 
BCS theory comes from the pairing of the nearly independent quasiparticles near Fermi energy. The effects 
of the interactions in $P_I$ on the superconductivity have been implicitly included in parameters of the 
renormalized quasiparticles. The adiabatic principle and the Pauli exclusion principle preserve the 
reliability of the weak-coupling BCS theory in good metal superconductors.

The strong-coupling Eliashberg theory is an updated version of the weak-coupling BCS theory. In the 
strong-coupling Eliashberg theory, the time retarded formation of the Cooper pairs and the scattering 
renormalization of the single-particle excitations are additionally included upon the weak-coupling 
approximations. While the weak-coupling BCS theory is a mean-field theory with static pairing potential, 
the strong-coupling Eliashberg theory can be regarded as a dynamical mean-field theory with dynamical 
pairing potential. In the latter theory, although the mother normal state is also the Fermi-liquid state, 
the physics of $P_{sc} P_{I}$ are treated unifiedly with the dynamical retarded pairing physics and the 
scattering renormalization of the single-particle excitations included simultaneously. However, unlike in 
the weak-coupling BCS theory where the ground and excited states can be given out clearly, in the 
strong-coupling Eliashberg theory the ground and excited states can not be directly shown, while only the 
physics of the particle and hole excitations are involved in a perturbative self-consistent theory with the 
dynamical pairing potential in an extended self-energy. 

Physically, both the weak-coupling BCS theory and the strong-coupling Eliashberg theory are in BCS formalisms 
where the pairing of the physical electrons is the essential concept for superconductivity. It should be noted 
that in the above two theoretical formalisms, only the inner-pair physics of the Cooper pairs is involved. 
The center-of-mass physics of the Cooper pairs is described by the macroscopic Ginzburg-Landau functional 
theory, where the detailed inner-pair physics as given in microscopic BCS formalisms is neglected. An updated 
formalism to unify both the microscopic pairing formation and the macroscopic pair condensate of the 
superconductivity can be established in a path integral method where both the single-particle and the pairing 
physics are treated unifiedly. The pairing fluctuations should be carefully included such as done in 
Reference~\cite{Nozieres1985}. 

\section*{References}


\end{document}